\documentclass[prl,twocolumn,aps,amssymb,footinbib]{revtex4}
\usepackage{amssymb}
\usepackage{graphicx}
\usepackage{amsmath}
\usepackage{times}
\usepackage{color}
\usepackage{subfigure}
\usepackage{setspace}
\usepackage{bm}% bold math

\begin{document}

\newcommand {\beq} {\begin{equation}}
\newcommand {\eeq} {\end{equation}}
\newcommand {\bqa} {\begin{eqnarray}}
\newcommand {\eqa} {\end{eqnarray}}
\newcommand {\da} {\ensuremath{d^\dagger}}
\newcommand {\ha} {\ensuremath{h^\dagger}}
\newcommand {\adag} {\ensuremath{a^\dagger}}
\newcommand {\no} {\nonumber}
\newcommand {\ep} {\ensuremath{\epsilon}}
\newcommand {\ca} {\ensuremath{c^\dagger}}
\newcommand {\ga} {\ensuremath{\gamma^\dagger}}
\newcommand {\gm} {\ensuremath{\gamma}}
\newcommand {\up} {\ensuremath{\uparrow}}
\newcommand {\dn} {\ensuremath{\downarrow}}
\newcommand {\ms} {\medskip}
\newcommand {\bs} {\bigskip}
\newcommand{\kk} {\ensuremath{{\bf k}}}
\newcommand{\rr} {\ensuremath{{\bf r}}}
\newcommand{\kp} {\ensuremath{{\bf k'}}}
\newcommand {\qq} {\ensuremath{{\bf q}}}
\newcommand{\nbr} {\ensuremath{\langle ij \rangle}}
\newcommand{\ncap} {\ensuremath{\hat{n}}}

\begin{abstract}

  We calculate the rate of creation of double occupancies in a 3D
  Fermionic Mott insulator near half-filling by modulation of optical
  lattice potential. At high temperatures, incoherent holes lead to a
  broad response peaked at the Hubbard repulsion $U$. At low
  temperatures, antiferromagnetic order leads to a coherent peak for
  the hole along with broad features representing spin wave shake-off
  processes. This is manifested in the doublon creation rate as a
  sharp absorption edge and oscillations as a function of modulating
  frequency. Thus, modulation spectroscopy can be used as a probe of
  antiferromagnetic order and nature of quasiparticle excitations in
  the system.

\end{abstract}

\title{Modulation Spectroscopy and Dynamics of Double 
Occupancies in a Fermionic Mott Insulator }

\author{  Rajdeep Sensarma, David Pekker, Mikhail D. Lukin and Eugene Demler}  
 \affiliation{ Physics Department,
Harvard University, Cambridge, Massachusetts 02138, USA}
\maketitle

\section*{}

Advances in experiments with cold atoms on optical lattices have made
them promising candidates for simulators of lattice models which play
an important part in our understanding of strongly interacting quantum
systems. Recently Mott insulating states\cite{Imada_rev} have been
obtained in the large $U$ limit of the repulsive Fermionic Hubbard
model\cite{fermi_hub,bloch_mott} with these systems.

% However, the field of cold atoms does not have the ready repertoire of
% experimental techniques available to condensed matter physicists to
% probe the many body physics in these systems; e.g. the spectral weight
% of charge (density) excitations are measured in typical material
% systems by optical conductivity measurements. However, the atoms in
% the optical lattice are electrically neutral, which renders this
% method ineffective
 
% Recently, experiments \cite{fermi_hub} have been performed where the
% modulation of the optical lattice potential is used to couple energy
% to the system. This leads to creation of double occupancies in the
% Mott insulators. The amount of double occupancies produced have been
% measured and a peak is observed in the response at a frequency
% matching the Hubbard repulsion $U$
% While the peak itself is not a signature of Mott
% insulator, the absence of response at low frequencies in the large
% $U/t$ limit can be taken as a signature of a Mott gap.  Under the
% circumstances, it is essential to understand the response
% theoretically and relate the measurements to quantities like spectral
% functions, which give us intuition about the many body physics in
% these strongly interacting systems.
%However, the field of cold atoms does not have the ready repertoire of
%well understood experimental techniques available to condensed matter
%physicists to probe many body physics. 
% It is important to
% understand what information about the strongly correlated physics one
% can obtain from this new experimental technique.

Motivated by recent experiments\cite{fermi_hub}, we give a theoretical
formulation of the response of a Mott insulator near half-filling to
modulation of optical lattice potential
%, focusing on the production of double occupancy in the
%system by the perturbation.  
by relating the rate of production of double occupancies to the
convolution of spectral function of holes and double occupancies
(doublons) in the Mott insulator.  We will show that this technique
can be used to detect the presence of antiferromagnetic (AF) order and
probe the nature of quasiparticle excitations (coherent
vs. incoherent) in the system. We also discuss the connection of this
response to optical conductivity in corresponding charged systems.

 %  At
% high temperature, the doublon and hole excitations are all incoherent.
% In the antiferromagnetic phase at low temperatures, a coherent peak
% appears on top of a series of incoherent peaks\cite{kane_lee,
%   ruck_prb}. This peak structure is reflected in the response as a
% sharp absorption edge followed by oscillations at higher
% frequencies. Therefore, lattice modulation spectroscopy can be used to
% observe the nature of quasiparticle excitations and thus detect the
% presence of antiferromagnetic ordering in the Mott insulating
% state. 
%% We use a Schwinger Boson Slave Fermion approach and relate the rate of
%% production of double occupancies (within perturbation theory) to the
%% convolution of spectral function of a single hole and a single double
%% occupancy (doublon) in the Mott insulator.
%We also discuss the connection of this response to optical conductivity 
%in corresponding charged systems. 

We focus on two temperature regimes: (i) the high
temperature limit ($J \ll T\sim t_h \ll U$), where $T$ is the
temperature, $t_h$ the tunneling matrix, and
$J=4t_h^2/U$ the super-exchange scale, which controls the quantum
dynamics of the background spins; and (ii) The low
temperature limit ($T \ll J$), with an AF ordered
spin background.

In the paramagnetic phase (current experiments\cite{fermi_hub}), we
get a response peaked at $\omega=U$ with a width equal to twice
the bandwidth of the holes, reflecting the completely incoherent
hole and doublon in this limit. We also
derive a sum rule for the energy integrated rate of doublon production
in this limit.

At low temperatures, the AF ordering leads to coherent propagation of 
quasiparticles and manifests itself in a sharp absorption edge in the 
production rate. Additional structures at higher energies appear as a 
result of shake-off processes of spin waves. Thus, lattice
modulation spectroscopy can be used to observe the nature of
quasiparticle excitations and detect the presence of
antiferromagnetic ordering in the Mott insulating state.

% In the low temperature limit, the AF ordered
% background leads to a coherent peak in the hole spectral function on
% top of a series of incoherent peaks which correspond to spin wave
% shake-off processes. The coherent part is reflected in the response as
% a peak at the lower end of the spectrum, resulting in a sharp
% absorption edge. The broad features due to shake-off processes lead to
% oscillations as a function of perturbing frequency.

{\bf Modulation of Optical Lattice:} Repulsive fermions in optical lattices 
are well described by a one band Hubbard model
\beq
H=-t_h\sum_{\nbr}\ca_{i\sigma}c_{j\sigma}+U\sum_i n_{i\up}n_{i\dn} 
\eeq   
where the tunneling matrix $t_h$ and the onsite repulsion $U$ depend on the 
depth of the optical lattice $V$ through \cite{deep_latt}
\beq
t_h\sim E_r\left(\frac{V}{E_r}\right)^{\frac{3}{4}}
e^{-2\sqrt{\frac{V}{E_r}}}, ~~~~U\sim\frac {a_s}{\lambda}
E_r\left(\frac{V}{E_r}\right)^{\frac{3}{4}}
\eeq
where $E_R$ is the recoil energy of the photon, $\lambda$ 
is its wavelength and 
$a_s$ is the s-wave scattering length of the atoms.

The modulation of the optical potential $V(t)=V_0+\delta V \sin
(\omega t)$ effects both $t_h$ and $U$. The modulation of $U$
can be neglected in the large $U/t_h$ limit. The modulation of
$t_h(t)=t_h+\delta t_h \sin(\omega t)$ is related to $V(t)$ by
\beq 
\delta t_h =t_h\delta V\left[\frac{3}{4V_0}-\frac{1}{\sqrt{V_0E_r}}\right] 
\eeq 
%Since tunneling can create double occupancies, the perturbation leads
%to production of double occupancies. However, 
The single particle spectrum of the Mott insulator is formed of two
bands: (a) the lower Hubbard band, which does not contain double
occupancies and is exactly filled at half-filling and (b) the upper
Hubbard band, which contains a single double occupancy and is
completely empty at half-filling. These are separated by the Mott gap
$\sim U$. So, the modulation of the optical barrier will produce
double occupancies once the frequency of modulation exceeds the Mott
gap.

{\bf Schwinger Bosons and Slave Fermions:} In the Schwinger Boson
Slave Fermion representation, we represent the singly occupied sites
(spins) by two Schwinger bosons $\adag_\sigma$, the doubly
occupied sites by a doublon $\da$ and the empty sites
by a holon $\ha$. The doublon and holon are Fermions. The original
Fermion creation operator can be written as $\ca_{i\sigma}=\adag_{i\sigma}h_i+\sigma
a_{i-\sigma}\da_i$ with the local constraint equation
$\adag_{i\sigma}a_{i\sigma}+\da_id_i+\ha_ih_i=1 $. 
\begin{figure}[floatfix]
\includegraphics[scale=0.3]{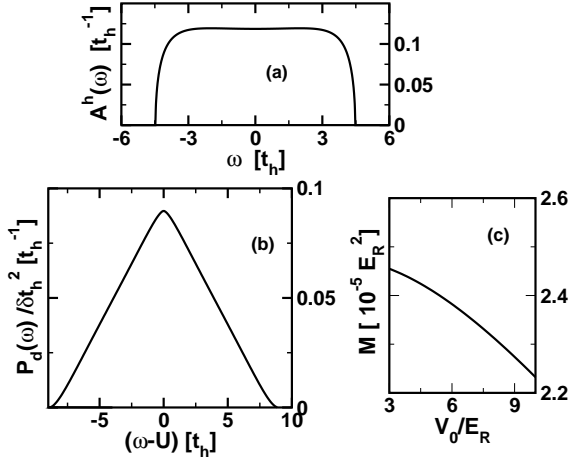}
\caption{(a) The density of states of a single hole in a half-filled background 
in the atomic limit. (b) The rate of production of double occupancies as a 
function of frequencies and (c) The energy integrated response in 
high temperature limit. }
\label{response}
\end{figure}
We define the following operators:
$F^\dagger_{ij}=\sum_\sigma\adag_{i\sigma}a_{j\sigma}$ and
$A^\dagger_{ij}=\sum_\sigma \sigma\adag_{i\sigma}\adag_{j-\sigma}$
which represent the hopping of the bosons and the creation of singlet
configurations. We further make the following
transformation: On B sublattice $\da \rightarrow -\da$. Then the
unperturbed Hamiltonian is
\bqa 
\no H_0=t_h\sum_{\nbr}(\da_id_j+\ha_ih_j)F_{ij}+(\da_i\ha_jA_{ij}+h.c.)\\
+U\sum_i \da_id_i 
\eqa 
while the perturbation due to the lattice
modulation is 
\beq 
H_1(t)=\delta t_h\sin[\omega \, t]\sum_{\nbr}\da_i\ha_jA_{ij}+h_id_jA^\dagger_{ij}
\eeq 
where we have neglected terms that do not create or destroy doublons. 
 We work with a system at half-filling, at
temperatures $T \ll U$, where we can neglect the presence of doublons
and holes in the unperturbed system. The number of doublons
created in time $t$ is given by 
\beq 
N_d(t)=\sum_n e^{-E_n/T}\langle
n| U^\dagger(t) \sum_i \da_i d_i U(t)|n\rangle 
\eeq 
where $|n\rangle$
denotes the unperturbed states (in our case spin configurations) with
energy $E_n$ and the time evolution operator has a perturbation
expansion $U(t)=1-i\int_0^t dt' H_1^I(t') -\int_0^t dt'
\int_0^{t'}dt'' H_1^I(t')H_1^I(t'')$, where the standard interaction
representation of an operator is given by ${\cal
  O}^I(t)=e^{iH_0t}{\cal O}(t)e^{-iH_0t}$.  We also assume that the
doublons are created by the action of the perturbation Hamiltonian
only, i.e. the time evolution of the system by the unperturbed
Hamiltonian conserves the number of doublons.  This neglects the decay
of doublons into a pair of bosons during the time evolution. This
approximation is justified as long as $T \ll U$, since, due to energy
conservation requirements, the decay of a doublon in the system is a
very slow process \cite{dbl_relax}. Under these assumptions, the first
order response of the system vanishes and upto second order in
perturbation theory, the rate of creation of doublons is
given by
\bqa
\no \displaystyle P_d(\omega)=\frac{\pi}{2}(\delta t_h)^2\int d\omega_1 \int d \omega_2 \sum_{\langle ij\rangle\langle lm\rangle}~~~~~~~~~~~~~~~~\\
 {\cal A}^s_{ijlm}(\omega_2){\cal A}^d_{il}(\omega_1){\cal A}^h_{jm}(\omega-\omega_1-\omega_2)
\eqa
where $\omega$ is the frequency of the perturbation, ${\cal A}^{d(h)}$
is the spectral function for the doublon (hole) and ${\cal A}^s$ is
the Fourier transform of $\langle A_{lm}(t)A^\dagger_{ij}(t')\rangle$.  
It is to be noted that we have
already used a mean field decoupling of the doublon, hole and
Schwinger boson operators to arrive at the above equation. This approximation 
is justified at large $U/t_h$ due to the separation of energy scales governing 
the hole (doublon) dynamics ($t_h$) and the spin dynamics ($J$). 

We emphasize that the response we are calculating is not 
equivalent to optical conductivity in the condensed matter systems. 
(i) The current vertex in optical 
conductivity is replaced by the kinetic energy vertex. (ii) The 
optical conductivity involves convolution of hole spectral function with 
itself, whereas the calculated response involves convolution of hole and 
doublon spectral functions. Since the doublon spectral function is shifted 
by $U$, as we move away from half-filling, there is no 
response at low frequencies, whereas there would be optical response 
at low frequencies in a compressible state.

{\bf High Temperature:} We now focus on the regime $U\gg T\sim t_h\gg J$, 
which is the regime of interest for the current
experiments. In this limit, the quantum dynamics of the spins are
irrelevant and one can replace the $A$ operators by the probability of
finding a $\up\dn$ or $\dn\up$ configuration in an ensemble where 
all spin configurations occur with equal weight. Thus,
\beq
P_d(\omega)=\frac{\pi}{2}P_s(\delta t_h)^2\sum_{\langle ij\rangle\langle lm\rangle}\int d\omega ' {\cal A}^d_{il}(\omega'){\cal A}^h_{jm}(\omega-\omega')
\eeq   
where $P_s$ is the probability of finding relevant configurations 
at $(i,j)$ and
$(l,m)$ given by $P_s=1/2$ if $(i,j)=(l,m)$, $P_s=1/8$ if $(i,j)$ and
$(l,m)$ have no overlap and $P_s=1/4$ if $(i,j)$ and $(l,m)$ share one
site in common. Due to particle-hole symmetry
of the problem at half-filing, we have ${\cal A}^d(\omega+U)={\cal
  A}^h(\omega)$, so that it is enough to compute the spectral function
for the holes only.

% In the following sections we will do two things: (a)  Calculate the spectral 
% function of a single hole (doublon) and hence the rate of production of 
% doublons due to shaking the lattice and (b) derive a sum rule for the 
%  energy integrated response of the system.

% \begin{figure}[t!]
% \begin{center}
% \includegraphics[scale=0.4]{feyn.pdf}
% \includegraphics[scale=0.4]{feyn_1.pdf}
% \includegraphics[scale=0.4]{feyn_2.pdf}
% \caption{The Feynman diagrams used to evaluate the spectral function of a 
% single hole. Solid lines indicate hole propagators and wavy lines indicate 
% boson propagators. Double solid lines indicate self-consistent propagators 
% for holes. (a) Interaction vertex between holon and Schwinger bosons (b) and 
% (c) Feynman diagrams used to calculate the holon Green's function. Note 
% that the contribution of diagram (b) vanishes due to a lattice momentum 
% summation}
% \end{center}
% \label{feyn}
% \end{figure}
%
 
We now try to evaluate the spectral function of a
single hole in a half-filled background where the
spins are completely disordered. The hole
is completely incoherent, i.e. it moves diffusively in the system. The
spectral function of the hole in this limit has been worked out by
Brinkman and Rice \cite{brinkman} and Kane {\it et al} \cite{kane_lee}
using the so called retraceable path approximation.
% \begin{figure}[t!]
% \begin{center}
% \includegraphics[scale=0.6]{integrated.pdf}
% \caption{ The frequency integrated rate of production of doublons as a function 
% of $V_0/E_R$ for modulation of optical lattice. A optical barrier modulation 
% of $\delta V_0/V_0$ of $0.1$ is used for the plot}
% \end{center}
% \label{srule}
% \end{figure}

The Green's function has contributions from processes where the hole hops 
from one point to another. However, as the hole hops, it scrambles the spin 
configuration and a string of ferromagnetic bonds is required along the path 
for the process to contribute. The probability 
of finding such a string is given by $(1/2)^L$, where $L$ is the 
length of the path. 
However, the trajectories where the hole retraces its path do not scramble 
the background spins and have a weight of $1$ as opposed to 
$(1/2)^L$. They provide the dominant contribution to the density of states 
at low energies.

The easiest way to derive the spectral function is to write the
Green's function as a function of the frequency $\omega$ in the
following way : $G^{-1}(\omega)=\omega[1-\Sigma(\omega)] $ and derive a
self-consistent equation for the self energy. The first contribution
to the self-energy comes from the hops to nearest neighbours and gives
$\Sigma^{(1)}(\omega)=zt_h^2/\omega^2$, where $z$ is the co-ordination
number of the lattice. To include the longer hops, the denominator of
$\Sigma^{(1)}$ can be modified with a higher order self-energy
$\Sigma^{(2)}(\omega)=zt_h^2/\omega^2(1-(z-1)t_h^2/\omega^2)$, where the
factor $z-1$ comes from excluding the initial site while considering
the initial hop. This method is similar to the one used by Anderson
in his original paper on localization physics \cite{And} and gives a
self energy 
$\Sigma(\omega)=(z/z-1)[1/2-\sqrt{\omega^2-4(z-1)t_h^2}/\omega]$.
This leads to the spectral function 
\beq 
{\cal
  A}(\omega)=\frac{1}{\pi z
  t_h}\left[\frac{(5-9\omega^2/z^2t_h^2)^{\frac{1}{2}}}{1-\omega^2/z^2t_h^2}\right]
\eeq 
The spectral function is plotted as a function of frequency in
Fig \ref{response}(a). The spectrum is incoherent and has a band-width
of $2\sqrt{z-1}t_h$. The spectral weight decreases monotonically as one
goes towards the band edge. The rate of production of doublons,
calculated using this spectral function is plotted in Fig \ref{response}(b). 
There is a peak around $\omega=U$ with weight upto twice the bandwidth 
(for the holes) around it. 

We note here some recent work in the paramagnetic phase using different
techniques\cite{huber,kollath}.

{\bf Sum Rule:} Sum-rules have played an important role in various
strongly correlated systems, since they often involve less
approximations and serve as useful check on both theory and
experiments. In this case we consider the energy integrated rate of
production of doublons $\int d\omega P_d(\omega)$. Using the identity
$\int_{-\infty}^\infty d\omega {\cal A}^{d(h)}_{ij}(\omega)=\delta_{ij}$
we obtain the sum rule in the high temperature limit
\beq
%\
M=\int_{-\infty}^\infty d\omega P_d(\omega)= \frac{\pi}{4}(\delta t_h)^2
\eeq 
The sum-rule is proportional to $(\delta t_h)^2$, which is proportional to 
$t_h^2$ for a constant fractional change in the amplitude of the lattice 
potential.
Assuming $U/t$ is tuned by tuning the lattice potential and a constant
fractional change in the amplitude of the potential ($\delta V_0/V_0$
is held fixed), one finds that in the Mott regime, the energy
integrated weight monotonically decreases with increase in $V_0/E_R$, 
as shown in Fig ~\ref{response} (c). 
 
{\bf Low Temperature (Antiferromagnetic phase) }: We now consider the
response of the system at $T=0$ in an AF ordered
phase. As we will see, in this regime, the spectral function of holes
has a sharp peak and a series of broad features. The sharp peak is
reflected in the doublon creation rate as a sharp absorption edge and
the broad features result in oscillations in the rate as a function of
modulating frequency.

This phase is characterized by Bose condensation of $\up$ and $\dn$
Schwinger Bosons on opposite (A and B) sublattices. In a $1/S$
expansion, the fluctuations are governed by the Holstein Primakoff
Hamiltonian
\beq
H=\sum_\kk \omega_\kk\alpha^\dagger_\kk\alpha_\kk
\eeq
where $\omega_\kk=zJ(1-\gamma_\kk^2)^{1/2}$ is the dispersion of the
spin wave with $\gamma_\kk=(1/3)[\cos k_x+\cos k_y +\cos k_z]$, and
the quasiparticle operators are given by $\alpha_\kk =u_\kk~
a_\kk-v_\kk~ a^\dagger_{-\kk}$ with
\beq 
a_\kk=\sum_{i\in A}a_{i\dn}e^{i\kk\cdot \rr_i}
+\sum_{j\in B}a_{j\up}e^{i\kk\cdot\rr_j}
\label{trans}
\eeq
The coherence factors $u_\kk$ and $v_\kk$ are given by
$u_\kk=(1/\sqrt{2})(\omega_\kk^{-1}+1)^{1/2}$ and
$v_\kk=-\text{sgn}(\gamma_\kk)(1/\sqrt{2})(\omega_\kk^{-1}-1)^{1/2}$.
\begin{figure}
\begin{center}
\includegraphics[scale=0.35,viewport=20 5 760 580, clip]{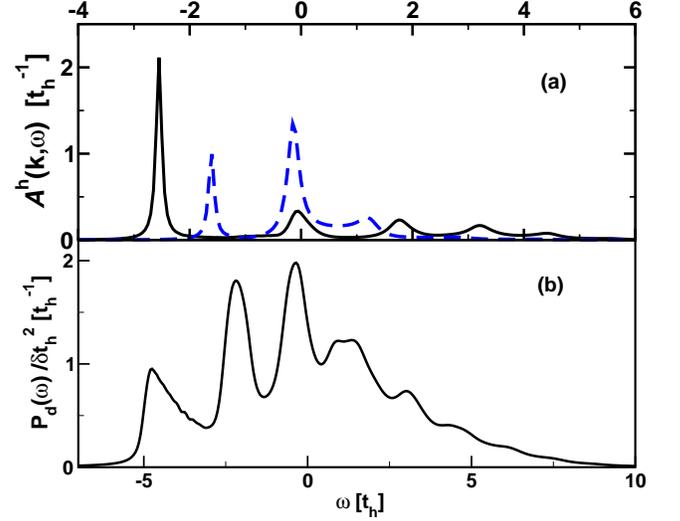}
\caption{(a)The spectral function of the hole at $(\pi/2,\pi/2,\pi/2)$ 
(solid line) and at $(0,0,0)$ (dashed line) for a system with 
$U/t=20$. The spectral functions are broadened by an artificial broadening 
of $0.06 t$ (b) The rate of doublon production $P_d(\omega)$ 
as a function of perturbing frequency.}
\label{akw}
\end{center}
\end{figure}
The hole hopping Hamiltonian can be written as
\beq
H^h=t_h\sum_{\langle ij\rangle}h^\dagger_ih_ja_{i\sigma}a^\dagger_{j\sigma}
\eeq
Replacing the $\up$ and $\dn$ spins on A and B sublattice by the condensate 
amplitude $\sqrt{\rho_0}=1$, this term can be written as 
\beq
H^h=zt_h\sqrt{\rho_0}\sum_{\kk\qq}h^\dagger_\kk h_{\kk-\qq}(u_\qq\gamma_{\kk-\qq}\alpha_\qq+v_\qq\gamma_\kk \alpha^\dagger_{-\qq}) + h.c.
\eeq
The motion of a hole is thus accompanied by creation of a spin
wave. We calculate the self-energy of the hole in a self-consistent
Born Approximation \cite{ruck_prb,ruck_prl} which is equivalent to 
calculating the non-crossing
Feynman diagram for the self energy. At $T=0$, the self-energy is given by
\beq 
\Sigma(\kk,\omega)=\sum_\qq|\Gamma(\kk,\qq)|^2G(\kk-\qq,\omega-\omega_\qq) 
\eeq
where the vertex function 
$\Gamma(\kk,\qq)=zt_h\sqrt{\rho_0}(u_\qq\gamma_{\kk-\qq}+v_\qq\gamma_\kk)$ and the 
self-consistency is ensured through 
\beq
G^{-1}(\kk,\omega)=\omega-\Sigma(\kk,\omega).
\eeq  
The spectral weight obtained from the self-consistent solution for
$J=0.2 t_h$ is plotted for two different $\kk$ values, ($[0,0,0]$ and
$[\pi/2,\pi/2,\pi/2]$) in Fig~\ref{akw}(a). At the lowest energy of
propagation of the hole, which occurs at $(\pi/2,\pi/2,\pi/2)$, spin
waves cannot be created (at $T=0$) and there is a coherent peak. The
coherent weight is largest at this point and gradually decreases as
one moves to the center of the Brillouin zone.  The location of the
coherent peak disperses as $\sim J \gamma_\kk^2$, corresponding to
second order hopping processes which do not scramble the
AF alignment. 

Beyond the coherent peak, there are
additional broad features at higher energies, whose peak to peak
distance scales with $J$. These are generated by spin wave shake-off
processes\cite{shraiman}.  The peaks correspond to
$2,4,6,...$ spin waves and are dominated by spin waves near the
Brillouin zone boundary where the flat spin-wave spectrum results in a
diverging density of states. This is similar to peaks in the 2-magnon
Raman response in antiferromagnetic insulators\cite{chubukov}.

In terms of the calculated spectral function, one can calculate the 
rate of doublon production as 
\beq
P_d(\omega-U)=\frac{\pi}{2}(\delta t_h)^2\kappa\sum_\kk \gamma_\kk^2\int d \omega_1 {\cal A}(\kk,\omega_1){\cal A}(\kk,\omega-\omega_1)
\eeq
where $\kappa=1-(1/2z)\sum_\kk \gamma_\kk^2/\omega_\kk$ is a vertex
correction which takes care of the singlet spectral function. 
There is no convolution with spin spectral functions 
as the spin dynamics is
governed by the transverse (phase) modes ($\up$ on B and $\dn$ on A
 sublattices) and the longitudinal (amplitude) modes ($\up$ on A
 and $\dn $ on B) of the condensate are neglected\cite{kane_lee}. 

 The rate of doublon production is plotted is
Fig~\ref{akw}(b). It shows an abrupt edge at the lower end of the spectrum
corresponding to the coherent spectral weight of the holes. The other
oscillations in the response reflect the convolution of the coherent
part with the broad incoherent peaks due to shake-off processes 
and that of the peaks
themselves. We thus see that the presence of the AF
order leaves its signature in the frequency dependence of the response.

We sketch what happens as we move away from half-filling 
(still remaining within the AF phase). The order parameter $\rho_0$ decreases, 
weakening the scattering of the hole by the spin waves. Thus the coherent 
part should grow leading to a sharper edge.
This is in contrast to disordering the AF phase by raising temperature, where the 
scattering from occupied spin wave modes reduce the coherent part. 

 Although our calculation is done for $T=0$, we expect these
qualitative features to be valid as long as the temperature is much
below the Neel transition temperature.
 
{\bf Comparison with Experiments:} In the experiments of
Ref.~\cite{fermi_hub}, the modulation is kept on for a fixed number of
cycles. The time of drive is proportional to $\omega^{-1}$ and the
quantity measured is proportional to $P_d(\omega)/\omega$. So the frequency
integrated response should decrease even faster with $U/t_h$ as
compared to Fig.~\ref{response}c. However, the experimental
data shows a monotonic increase with $U/t_h$\cite{ess_expt}. 

We believe this could be due to several reasons: (a) The
response might be dominated by terms beyond second order perturbation
theory. This can be checked by putting on the drive for different amount 
of time and looking at the linearity (or lack thereof) of the number of 
doublons produced with time. 
(b) The unperturbed system is not in thermal equilibrium due
to slow relaxation of the doublons created during tuning $U/t$
\cite{dbl_relax} or (c) Relaxation of doublons while driving the 
system leads to a steady state behaviour.
We hope the discrepancies between the theory and experiments can be settled 
with further experiments on this system. 

{\bf Conclusion:} We have related the rate of production of double
occupancies by modulation of optical lattice in a 3D Mott insulator
near half-filling to the convolution of hole and doublon spectral
functions. This technique can be used to study the nature of 
quasiparticle excitations and detect presence of AF order in the 
system. In the paramagnetic phase there is a broad response 
peaked around $\omega=U$. In the ordered phase, the coherent hole
is reflected as a sharp absorption edge, while shake off processes lead 
to oscillations.

% At low temperatures the antiferromagnetic ordering leads to
% a coherent peak in both the hole and doublon spectral functions along
% with broad features representing shake-off processes. The coherent
% part shows up in the response as a sharp absorption edge, while the
% broad features lead to oscillations. This probe can thus be used to 
% detect the presence of antiferromagnetism in the system.
% We have analyzed the creation of double occupancies in a 
% 3D Fermionic Mott insulator near half filling due to optical lattice 
% modulations in $2^{nd}$ order perturbation theory. Using a Schwinger Boson 
% Slave Fermion representation, we have related the rate of doublon production 
% to a convolution of the hole and doublon spectral functions. At high 
% temperature, we find a broad peak at $\omega=U$ reflecting the incoherent 
% nature of the holes. We derive a sum rule for the frequency integrated response 
% in this limit. At low temperatures the antiferromagnetic ordering 
% leads to a coherent peak in the hole spectral function along with broad 
% features representing shake-off processes. The coherent part shows up in the 
% response as a sharp absorption edge, while the broad features lead to 
% oscillations. We hope future experiments in 
% the anti-ferromagnetic state will observe this 
%features.  

We would like to thank Ehud Altman and Sebastian Huber for useful discussions. 
This authors acknowledge the support of DARPA, CUA and AFPOSR. R.S. and D.P. 
was partially supported by NSF grant DMR-05-41988.


\begin{thebibliography}{99}
\bibitem{deep_latt} This relation holds for deep
  lattices, which is the case for the Mott insulating regime we are interested
  in.
\end{thebibliography}
\end{document}